\documentclass[aps,prl,twocolumn]{revtex4}
\usepackage{amsmath,bm,epsfig}
\usepackage{amssymb}
\begin{document}

\title{Comparison of Static Length-Scales Characterizing the Glass
Transition}
\author{Giulio Biroli$^1$, Smarajit Karmakar$^2$ and Itamar Procaccia$^3$}
\affiliation{$^1$IPhT, CEA/DSM-CNRS/URA 2306, CEA Saclay, F-91191 Gif-sur-Yvette Cedex, France}
\affiliation{$^2$TIFR Centre for Interdisciplinary Sciences, Tata Institute of Fundamental Research,
21 Brundavan Colony, Narsingi, Hyderabad - 500075}
\affiliation{$^3$Dept. of Chemical Physics, The Weizmann Institute of Science, Rehovot 76100, Israel}


\begin{abstract}
The dramatic dynamic slowing down associated with the glass transition is considered by many to be related to the existence of a
static length scale that grows when temperature decreases. Defining, identifying and measuring such a length is a subtle and
non-trivial problem. Recently, two proposals, based on very different insights regarding the relevant physics, were put forward.
One approach is based on the point-to-set correlation technique and the other on the scale where the lowest eigenvalue of the
Hessian matrix becomes sensitive to disorder. In this Letter we present numerical evidence that the two approaches might result
in the same identical length scale. This provides further mutual support to their relevance and, at the same time, raise
interesting theoretical questions, discussed in the conclusion, concerning the fundamental reason for their relationship.

\end{abstract}
\maketitle
%%%%%%%%%%%%%%%%%%%%%%%%%%%%%%%%%%

A natural assumption that prevails in the physics community regarding the glass transition is that together with the spectacular
slowing down associated with this phenomenon there should be also a length scale that increases in accordance with the increased
time scale \cite{11BB}. Decades of research efforts were devoted to find such a length \cite{01Donth, 01DS}. The discovery of
dynamic heterogeneity  in super-cooled liquids both in experiments and in theoretical studies ~\cite{BookDH} and the detailed
analysis of the associated dynamical length-scale using multi-point correlation
functions~\cite{95HH,99BDBG,02DFGP,00FP,05Chi4,PREchi4,09KDS,10KDS} revealed that more and more particles move in a correlated
way approaching the glass transition. There is no doubt by now that there exists a growing {\it dynamic} length accompanying
the glass transition. Although this is a new and important facet of the phenomenology of super-cooled liquids, it is still not
clear to what extent dynamic correlations are the consequence or the primary origin of slow dynamics~\cite{09KDS}.
Another natural candidate for the dominant length-scale of glassy dynamics is a {\it static}
length. Finding it remained an open problem for a long time. Recently, two very specific and very different methods to determine
the elusive characteristic length scales had been proposed and implemented. In this Letter we review both methods and provide
numerical indications that although the two methods are extremely different, they seem to lead to the same length scale. This
provides further mutual support to their relevance and, at the same time, raises interesting theoretical questions concerning
the fundamental reason for their relationship.

The first method is based on the "point-to-set" (PTS) length \cite{04BB,MM} that will denote henceforth as $\xi_{_{\rm PTS}}$. This
length, originally introduced to characterize the real space structure of the so called ``mosaic state'' arising in the Random
First Order Transition theory \cite{04BB,reviewRFOT}, allows one to probe the spatial extent of positional amorphous order. It
has attracted a lot of attention recently. In particular, it was measured in several numerical simulations
~\cite{CGV,08BBCGV,SauTar10,BeKo_PS,HocRei12} and shown to grow mildly in the (rather high) temperature regime investigated.
Rigorous results have also strengthened its relevance: in \cite{SM} it was proven that if the relaxation time-scale, $\tau_\alpha$,
diverges in a super-Arrhenius way, either at finite temperature or at zero temperature, then $\xi_{_{\rm PTS}}$ has to diverge too,
at least as fast as $(T\log \tau)^{1/d}$ ($d$ being the spatial dimension). The definition of the point-to-set length is the
following: take a typical equilibrium configuration, freeze the positions of all particles outside a sphere centered around a
given point and study how the thermodynamics of the remaining particles, inside the sphere, is influenced by this amorphous
boundary condition; $\xi_{_{\rm PTS}}$ is the smallest radius of the sphere at which the boundary has no longer any effect
on the configuration at the center. This procedure is very similar to the one used to demonstrate the existence of long-range
order for, e.g., the 3D Ising model: setting the boundary spins up at low temperature forces all configurations, even far from
the boundary, to be in the up-state. The crucial difficulty in supercooled liquids is that we do not know apriori what is the
correct boundary condition favoring amorphous order; the trick of freezing the particle positions outside the cavity uses the
fact that instead the system ``does'' it, if it is indeed becoming more and more spatially statically correlated and amorphously
ordered. In practical implementation, an overlap function (defined below) is used to estimate the degree of similarity in the
center of the cavity between the initial configuration and the one equilibrated in presence of the frozen amorphous boundary
conditions.

The overlap function used for this study is same as in \cite{HocRei12} and is given by
\begin{equation}
q(R) = \lim_{t\to\infty}\frac{1}{N_{\nu}l^3}\sum_{i=1,N_{\nu}} \langle n_i(t) n_i(0) \rangle
\end{equation}
where $N_{\nu}$ is the number of cubic boxes considered in the center of the cavity with
$l^3 = 0.050$ being the volume of the individual cubes; $n_i(t)$ is the number density of particles
inside the cube $i$ at time $t$ and the angular average implies both the thermal averaging as well as
averaging over the frozen boundary condition. This overlap is defined in such a way that for two
identical configurations the overlap is unity and for two completely uncorrelated configurations
it will be $q_c = l^3\rho = 0.060$, where $\rho$ is the number density of particles.
We have implemented the Particle Swap Annealing (PSA) method (see
Ref.\cite{HocRei12} for details ) with Molecular Dynamics simulation to calculate the overlap
correlation function for different radius $R$ of the cavity. The length scale was then estimated by
fitting the overlap function for different $R$ by the fitting function
$q(R)-q_c = A\exp\left[-\left(\frac{R - 1.0}{\xi}\right)^{\eta}\right]$ with $A$ as fixed
number for all temperatures. This choice of $A$ is partially motivated by the work in Ref.\cite{HocRei12}
and also to reduce the number of parameters to fit. In our case
we found that $A = 0.465$ fits the data better. It is important to note that fitting with
three free parameters give somewhat different estimation for the length scale but the quality of the
fit is very similar to that of the two parameters fit. As the number of data points are somewhat small
because of the extensive computational effort needed to calculate these overlaps for different
cavity radius, we decided to stick to the two parameters fit as done in \cite{HocRei12} for extracting
the point-to-set length scale.

Another, superficially unrelated way to define a static scale was announced in Ref. \cite{11KLP} and employed
further in \cite{12KP}. The starting point of the second method is the fact that
at low frequency the tail of the density of state (DOS) of amorphous solids consisting of $N$ particles
reflects the excess of plastic modes which do not exist in the density of states of purely elastic solid
\cite{02TWLB,10Sok}. This excess of modes is sometime referred to as the `Boson Peak' \cite{09IPRS}.
Here and below the 'mode' refers to the eigenfunction of the underlying Hessian matrix calculated at the
local minimum of the potential energy function $U$. The Hessian matrix is defined as
${\cal H}_{ij}^{\alpha \beta} = \frac{\partial^2 U}{\partial x_i^{\alpha} \partial x_j^{\beta}}$ where
$x_{i}^{\alpha}$ denotes the $\alpha^{th}$ component of coordinate of particle $i$. Recently \cite{11HKLP}
it was proposed that the eigenvalues $\{\lambda_i\}_{i=1}^{dN}$ (with $d$ being the space dimension) appear
in two distinct families in generic amorphous solids, one corresponding
to eigenvalues of the Hessian matrix that are only weakly sensitive to external strains; the other group consists of
eigenvalues that go to zero at certain values of the external strain, thus leading to a plastic failure. The first
group of modes is decently described by the Debye model of an elastic body, but this is not the case for the
second group  corresponding to the density of plastic modes.

To clarify this distinction further
write the excess part (in the thermodynamic limit) as $B(T)f_{\rm pl}\left(\frac{\lambda}{ \lambda_D }\right)$,
where the pre-factor $B(T)$ depends on the temperature and $\lambda_D \simeq \mu \rho^{2/d - 1}$ is the Debye
cutoff frequency and $\mu$ is the shear modulus. Particular models for the function
$f_{\rm pl}\left(\frac{\lambda}{ \lambda_D }\right)$ were presented
in \cite{10KLP}. We do not need to specify a function here, and it is only important to understand that this function
is a partial characterization of the degree of disorder which decreases upon approaching the glass transition.
Together with the standard Debye contribution one can approximate the low-frequency tail of the density of
states as
\begin{equation}
P(\lambda) \simeq A\left( \frac{\lambda}{ \lambda_D } \right)^{\frac{d-2}{2}}
+ B(T) f_{\rm pl}\left( \frac{\lambda}{ \lambda_D } \right) \ . \label{Poflam}
\end{equation}

The physical idea that allows the determination of the static typical scale is that the {\em minimal} eigenvalue
$\lambda_{\rm min}$ observed in a system of $N$ particles will be determined by either the first {\em or}
the second term in Eq. (\ref{Poflam}). For a system large enough, local disorder will be irrelevant in determining
$\lambda_{\rm min}$, and it will be determined by the Debye contribution. For small systems the opposite is true.
Thus there exists a system size where a cross-over occurs. This cross-over is interpreted
in terms of a typical length-scale separating correlated disorder from asymptotic elasticity. The length scale
computed in this way is denoted below as $\xi_{_{\lambda}}$.  One crucial element of the algorithm is that we need
to take a supercooled liquid at temperature $T$ and quench it rapidly to $T=0$ where the Hessian matrix is evaluated.
Since every time we hit a different "inherent state" we must consider the average $\langle \lambda_{\rm min}\rangle$
over many realizations. The full details of the algorithm of how to extract this length
can be found in Ref. \cite{11KLP}.

%%%%%%%%%%%%%%%%%%%%%%%%%%%%%%%%%%%
\begin{figure}
\centerline{\includegraphics[scale = 0.400]{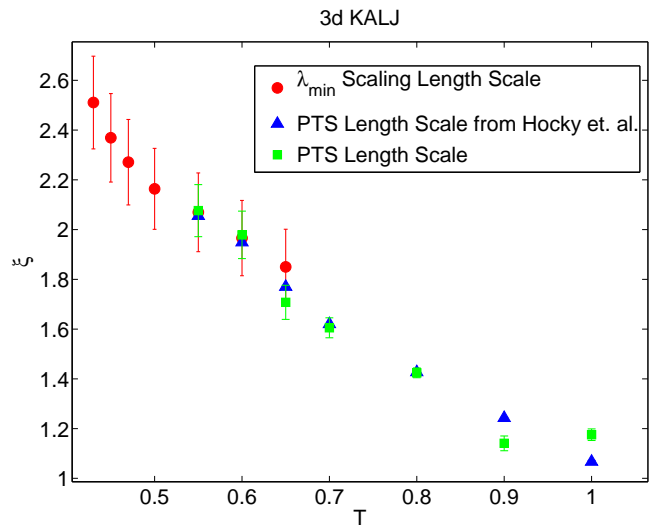}}
\caption{Comparison of the two length scales $\xi_{_{\rm PTS}}$ and $\xi_{_{\lambda}}$ for the 3-dimensional Kob
Andersen model. Also shows the comparison of the point-to-set length scale with that from Hocky {\it et. al.} Ref.\cite{HocRei12}.}
\label{compare1}
\end{figure}
%%%%%%%%%%%%%%%%%%%%%%%%%%%%%%%%%%%
\begin{figure}
\centerline{\includegraphics[scale = 0.43]{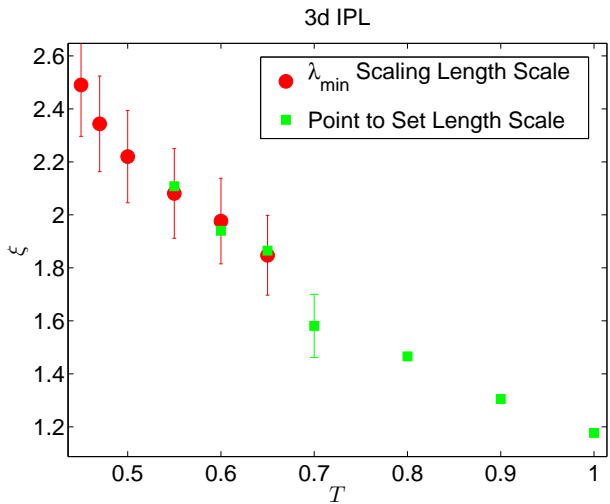}}
\caption{Comparison of the two length scales $\xi_{_{\rm PTS}}$ and $\xi_{_{\lambda}}$ for the 3-dimensional
inverse power law potential. Data for point-to-set length scale is taken from Ref.\cite{HocRei12}.}
\label{compare2}
\end{figure}
%%%%%%%%%%%%%%%%%%%%%%%%%%%%%%%%%%%%%%%%%%%%%%%%

Next we present evidence that suggests that the two length scales might be in fact the same. To do so we need to
compute them for the same glass formers over a range of temperatures that has at least a region of overlap.
It should be noted that the point to set method is easier to perform at higher temperatures, since we need to
equilibrate the system in the radial cavity, and this takes longer and longer at lower temperature. On the
other hand obtaining  $\xi_{\lambda}$ is easier at lower temperatures, since the information obtained from
the Hessian matrix becomes less relevant at high temperatures due to non-harmonic effects. Thus the two methods
are actually complementary. The results for the two length scales are shown for two popular models of
super-cooled liquids in Figs. \ref{compare1} and \ref{compare2}. The two models are (i) the standard
3-dimensional Kob-Andersen binary Lennard-Jones mixture (3d KALJ) \cite{95KA} and (ii) a 3-dimensional system
characterized by a repulsive inverse power law potential (3d IPL) as constructed in Ref. \cite{10PSD}.

The simulations
are done in the NVT ensemble for number density $\rho = 1.20$ for both these model systems. The details of the
parameters of the potentials can be found in \cite{95KA} for the Kob-Andersen model and in \cite{10PSD} for the
inverse power law model. We have done the simulations for system sizes ranging from $N = 150$ to $N = 10000$
to perform the finite size scaling analysis of $\lambda_{\rm min}$ in the temperature range $T \in [0.650,\, 0.430]$.
$\lambda_{\rm min}$ for each system size and temperature is averaged over $2000$ inherent structures. On the other
hand for the calculation of the PTS length scale, the simulations are done for $N = 4000$ system size for the temperatures
in the range $T \in [1.00, \, 0.550]$ and then overlaps are calculated for the cavities of radii in the range
$R \in [2.00,\, 5.00]$. To reach the asymptotic value of the overlap function for a given radius at some temperature, simulations are
performed for $2 \times 10^7$ molecular dynamics steps with particle swap annealing (PSA) (see supplementary materials
of Ref.\cite{HocRei12} for the details of the PSA method). The overlap is then averaged over $20$ different realizations
of the frozen boundary. In Fig.\ref{compare1}, one can see that our estimation for the PTS
length scale for the 3d KALJ model matches very well with that of \cite{HocRei12}. We took the estimation of PTS length
scale for the  3d IPL  model from Ref.\cite{HocRei12}.

Note that it is quite impossible at present to compute $\xi_{_{\rm PTS}}$ for lower temperatures in any one of the two models.
It is also difficult to compute $\xi_{_{\lambda}}$ for higher temperatures than those shown. Thus the two approaches
complement each other quite nicely. Our results, see Fig. 1 and 2, suggest that there the two length scales can be actually
the same: $\xi=\xi_{_{\rm PTS}}=\xi_{_{\lambda}}$. As the length scale from the finite size scaling of the minimum
eigenvalue can be determined up to a constant pre-factor (same for all temperatures), we rescaled the length scale
obtained from this method appropriately to match with that of the point-to-set length scale.
One should appreciate the fact that with all the efforts spent, the increase in length scale observed here,
combining the two techniques, is only by a factor of $2.5$. At the same time the relaxation time in both models
increases, for the same range of temperatures, of some $4$ orders of magnitude. It is thus clear that one can run
out of computer time to equilibrate our systems at lower temperatures to achieve a trustable estimate of the
length scale, and a new approach for reaching lower temperatures must be found.

%But already at this point one cam make a few comments of some relevance to the general theory of the glass transition. First we note, as seen in Figs. \ref{fit1} and \ref{fit2}, that the data obtained for the two models can be fitted to a ``law"
%\begin{equation}
%\xi =A+\frac{B}{\sqrt{T}} \quad \text{cf. Figs. \ref{fit1} and \ref{fit2}} \ . \label{fit}
%\end{equation}
%%%%%%%%%%%%%%%%%%%%%%%%%%%%%%%%%%%
%\begin{figure}
%\includegraphics[scale = 0.35]{GiulioFig3.eps}
%\caption{Fit of the length scale of the 3-dimensional Kob-Andersen model the ``'law" Eq. (\ref{fit}).}
%\label{fit1}
%\end{figure}
%%%%%%%%%%%%%%%%%%%%%%%%%%%%%%%%%%%
%\begin{figure}
%\includegraphics[scale = 0.40]{GiulioFig4.eps}
%\caption{Fit of the length scale of the 3-dimensional Inverse Power Law model to the ``'law" Eq. (\ref{fit})}.
%\label{fit2}
%\end{figure}
%%%%%%%%%%%%%%%%%%%%%%%%%%%%%%%%%%%%%%%%%%%%%%%%
%If we accept this fit at face value, it indicates that there is no finite temperature singularity associated
%with the glass transition, nor is there a finite Kauzman temperature. There may be however a zero temperature singularity. If this is the case the Kauzman temperature would be $T=0$ without any contradiction with the Third Law of thermodynamics. We stress however that it is still premature to state such strong statements on the basis of the
%present fit, we need to find a way to go to much lower temperatures before we could justify an extrapolation
%to $T\to 0$.

As discussed in the introduction, the interest in unveiling growing length-scales in super-cooled liquids is
due to their possible relationships with the growing time-scale. The main problem is to identify the length
that is at the root of the dynamical process inducing the slow dynamics, not just a length that is
correlated to it (which is almost tautologically the case
if they both grow with decreasing temperature). In theoretical models, from Adam-Gibbs \cite{65AG} to
RFOT \cite{reviewRFOT} and also in other complementary approaches 
the static length, in particular the point-to-set, is thought to be related to the spatial extent
of the dynamically activated cooperative processes leading to relaxation and flow. In consequence,
it is assumed that the  relationship between time and length is
\begin{equation}\label{fit}
\tau_\alpha = \tau_0 \exp(\Delta \xi^\psi/T)\ ,
\end{equation}
with $\psi$ being an a-priori unknown exponent of the order of unity, and $\tau_0$ and $\Delta$ being fitting parameters.
Although the variation of the static length we measured is very modest it is nevertheless interesting to test this
law with our present data. The relaxation time was computed for both models at all the temperatures for which
we have measured $\xi$ as follows. We first equilibrate a system of $N = 4000$ particles at a chosen temperature $T$.
Next we calculate the overlap function defined below to estimate the structural relaxation time $\tau_{\alpha}$,
\begin{equation}
Q(t) = \frac{1}{N}\sum_{i=1}^N w(|\vec{r}_i(t) - \vec{r}_i(0)|),
\end{equation}
where the weight function $w(x) = 1$ if $x<0.30$ and zero otherwise.
The overlap function $Q(t)$ is averaged over $20$ different initial conditions, and
the relaxation time $\tau_{\alpha}$ is defined to be the time where the correlation function $Q(t) = 1/e$,
where $e$ is the base of natural logarithm.

\begin{figure}
\includegraphics[angle=-90,scale = 0.30]{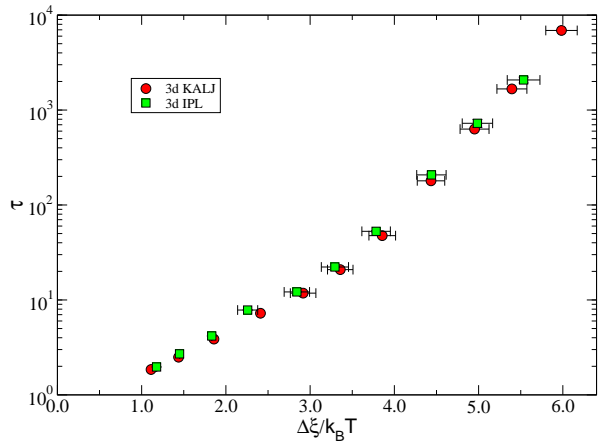}
\caption{Relaxation time $\tau_{\alpha}$ plotted as a function of $\xi/K_BT$ for two different model systems.
The pre-factor $\Delta$ is chosen appropriately to bring these two data set fall on top of each other. The values
of $\Delta$ for these two model systems are order of unity. }
\label{tau1}
\end{figure}

\begin{figure}
\includegraphics[angle=-90,scale = 0.30]{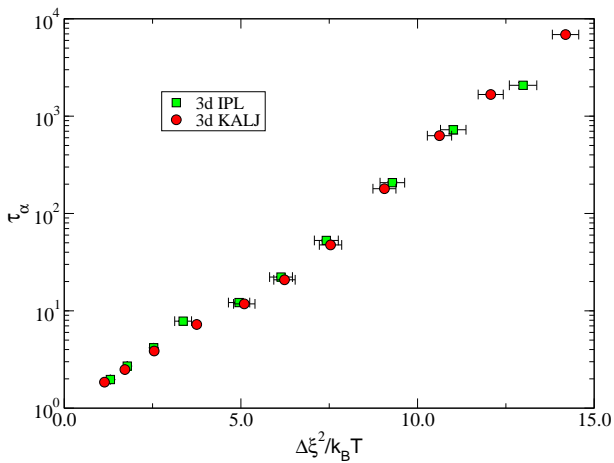}
\caption{Similar plot as that of Fig.\ref{tau1}, but this time relaxation time is plotted as a function of $\xi^2/K_BT$. The resulting
straight line indicates that the exponent $\psi$ is probably close to $2$ (see text for details).}
\label{tau2}
\end{figure}

In Figs.~\ref{tau1} and \ref{tau2} we show the dependence of $\ln (\tau_{\alpha})$
as a function of $\Delta \xi^\psi/T$  for the two models discussed above, with $\psi=1$ and $\psi=2$.
%%%%%%%%%%%%%%%%%%%%%%%%%%%%%%%%%%%
%%%%%%%%%%%%%%%%%%%%%%%%%%%%%%%%%%%
%%%%%%%%%%%%%%%%%%%%%%%%%%%%%%%%%%%%%%%%%%%%%%%%
From these results it seems that the linear fit of $\ln\tau$ with $\xi/T$ is very good once we focus on the super-cooled
regime where the relaxation time is of the order of $10$ or more, as also found in \cite{HocRei12,cavagnacammarota}. On the
other hand, if we want to have the whole temperature fit well represented then the quadratic fit of $\ln\tau$ with $\xi^2/T$
might be better. The exponent advocated by Kirkpatrick, Thirumalai and Wolynes for RFOT instead is $\psi=3/2$ \cite{reviewRFOT}.
Clearly, the jury must remain out until a larger range of temperatures becomes available.

In summary, we have compared the static lengths obtained studying point-to-set correlations and low frequency vibration modes in
two different models of super-cooled liquids. The most natural interpretation of our data is that these two lengths are
in fact identical (up to a rescaling factor). In a simple critical phenomenon, where below
the upper critical dimension a simple scaling theory based on a unique length $\xi$ rules the physical behavior, it would
be normal to expect that all relevant lengths are proportional one to the other. The glass transition is certainly a more
complicated phenomenon, already known to display different length-scales, e.g. static and dynamic ones \cite{andalo}.
In consequence, the (apparent) coincidence between the two length-scales observed in this work deserves a thorough physical
explanation. The point-to-set length measures the spatial extent of amorphous order: beyond $\xi_{_{\rm PTS}}$ a liquid
configuration is formed by statically independent patches, whereas below it particle positions are subtly correlated.
Put it in another way, below $\xi_{_{\rm PTS}}$ the system is amorphously ordered---it is an ideal glass---above it is a liquid.
If $\xi_{_{\rm PTS}}$ were indeed of the same order of $\xi_\lambda$ down to $T_g$ (and even below it) then this would tell
us that the low frequency modes of vibration of an ideal glass are very different from the Debye plane-waves characteristic
of crystalline materials. This brings about important practical and physical consequences: the computation of $\xi_{_{\rm PTS}}$
at low temperature could be replaced by the much more easier computation of $\xi_\lambda$ which just needs an equilibrated
configurations to start with. More interesting physically, vibrational properties of 
ultra-low temperature glasses would be related to the spatial extent of amorphous order reached at $T_g$; this would remain frozen in and would determine low frequency vibrational modes for $T\ll T_g$, as proposed in \cite{lubchenkowolynes}. 
What is the physical explanation for the existence of these
anomalous modes and whether these could possibly be related to a possible full replica symmetry breaking structure of
low temperature glasses \cite{parisiurbani} are open (and for the moment quite unclear) questions that deserve further work.

This work had been supported in part by an ERC ``ideas" grant STANPAS, an ERC grant NPRGGLASS, the Israel Science Foundation
and by the German Israeli Foundation.

\end{document}